\documentclass[submission,copyright,creativecommons]{eptcs}
\providecommand{\event}{WORDS 2011} 
\usepackage{breakurl}             

\usepackage{graphicx,eepic}
\usepackage{amsmath,amsfonts,amssymb}

\title{Optimizing Properties of Balanced Words}
\author{Nikita Sidorov
\institute{School of Mathematics\\
The University of Manchester\\ Oxford Road\\
Manchester M13 9PL\\ United Kingdom}
\email{sidorov@manchester.ac.uk}
}
\def\titlerunning{Optimizing Properties of Balanced Words}
\def\authorrunning{Nikita Sidorov}

\newcommand{\be}{\beta}

\newcommand{\al}{\alpha}
\newcommand{\ga}{\gamma}

\newcommand{\M}{\mathcal M}

\newtheorem{lemma}{Lemma}

\newtheorem{thm}[lemma]{Theorem}

\newtheorem{example}[lemma]{Example}

\providecommand{\urlalt}[2]{\href{#1}{#2}} \providecommand{\doi}[1]{doi:\urlalt{http://dx.doi.org/#1}{#1}}

\begin{document}
\maketitle

\begin{abstract}
In the past few decades there has been a good deal of papers which are concerned with optimization problems in different areas of mathematics (along 0-1 words, finite or infinite) and which yield -- sometimes quite unexpectedly -- balanced words as optimal. In this note we list some key results along these lines known to date.

\end{abstract}

Firstly, we recall that a finite or infinite 0-1 word $w=w_1w_2\dots$ is called \emph{balanced} if for every pair of finite subwords $u,v$ such that $|u|=|v|$, we necessarily have $||u|_1-|v|_1| \leq 1$, where $|u|_1=\#\{j: u_j=1\}$ stands for the 1-{\em length} of $u$. An infinite balanced word which is not eventually periodic is called \emph{Sturmian}.

There are several equivalent definitions of Sturmian sequences. Let $w=w_1w_2\dots$ be an infinite 0-1 sequence and put $p_w(n)=\#\{w_j\dots w_{j+n-1} : j\ge1\}$ -- the \emph{complexity function} of $w$. Then $w$ is Sturmian if and only if $p_w(n)=n+1$. The 1-{\em ratio} $\gamma=\lim_{n\to\infty} |w_1\dots w_n|_1/n$ is well defined for any Sturmian sequence $w$; furthermore, any Sturmian sequence with the 1-ratio $\gamma$ can be obtained by the formula
\begin{equation}\label{eq:sturmian}
w_n=\lfloor (n+1)\gamma +\delta\rfloor - \lfloor n\gamma +\delta\rfloor
\end{equation}
for some $\delta\in[0,1)$. For more details see, e.g., \cite{Loth}.

This survey paper is concerned with some optimization problems from various areas of mathematics and physics, in which balanced words (and, in some cases, Sturmian sequences) turn out to be optimizing.

\section{Multimodular functions and queuing}

We will begin with optimization problems in mathematics. Our first example comes from the seminal paper \cite{Ha}.

Define vectors $f_0, f_1, \dots, f_m$ in $\mathbb Z^m$ as follows: $f_0=(-1,0,\dots, 0,0), f_1=(1,-1,0,0,\dots, 0), f_2=(0,1,-1,0,\dots, 0), \dots, f_m=(0,0,\dots,0,1)$. Let now $\mathcal F=\{f_0,\dots, f_m\}$. We say that a function $J:\mathbb Z^m\to \mathbb R$ is {\em multimodular} if for any $u\in\mathbb Z^m$ we have \[
J(u+v)+J(u+w)\ge J(u)+J(u+v+w)
\]
for all $v,w\in\mathcal F$ with $v\neq w$. The function
\[
\underline{J}(z)=\sup\,\{A(z) : A\ \text{is affine and}\ A(u)\le J(u)\ \ \text{for all}\ \ u\in\mathbb Z^m\}
\]
is called the {\em lower convex envelope} of $J$.

\begin{thm} [\cite{Ha}]
Let $J$ be a multimodular function on $\mathbb Z^m$ and let $\underline{J}$ denote its lower convex envelope. If $x$ is any infinite 0-1 sequence with the 1-ratio $\gamma$, then
\[
\liminf_{n\to\infty} \frac1n\sum_{k=1}^n J(x_k,x_{k+1},\dots, x_{k+m-1})\ge \underline J(\widehat \gamma),
\]
where $\widehat\gamma=(\gamma,\dots,\gamma)$. Moreover, if $x=w$ given by (\ref{eq:sturmian}) for some $\delta$, then we have the equality.
\end{thm}

The author then applies this result to the following queuing problem: consider a sequence of customers arriving at a fixed rate in such a way that the interarrival times are i.i.d. Poisson random variables with the same finite mean. A 0-1 input sequence $x=(x_1,x_2,\dots)$ determines what happens to the $k$th customer, namely, if $x_k=1$, the customer is admitted and if $x_k=0$, he is sent elsewhere. The mean service time is assumed to be fixed. The number in the queue after customer~$k$ arrives, is $N_k+x_k$. Then the expectation of $\max N_k$ is a multimodular function.

Consequently, it follows from the above theorem that if a fraction $\gamma$ of customers is sent to a server queue according to a splitting sequence $x$, then the long-term average is minimized when $x_k=\lfloor (n+1)\gamma \rfloor - \lfloor n\gamma \rfloor$, i.e., along Sturmian sequences.

\section{Cyclic permutations of binary expansions}

Let $w=w_1\dots w_m$ be a finite 0-1 word and put
\[
b(w)=\sum_{k=1}^m w_k2^{m-k}.
\]
Now define
\[
B(w)=b(w)\cdot b(w^{(2)})\cdot b(w^{(m)}),
\]
where $w^{(1)}=w, w^{(2)}, \dots, w^{(m)}$ are the cyclic permutations of $w$. Let now $\mathcal W_{p,q}$ denote the set of 0-1 words of length $q$ with the 1-length $p$. As is well known (see, e.g., \cite{Loth}), there are precisely $q$ balanced words in $\mathcal W_{p,q}$, all of which are in the same {\em orbit} ($=$ all cyclic permutations of a word), so if $\mathbb W_{p,q}$ is defined to be the set of all orbits of words in $\mathcal W_{p,q}$, there is a unique balanced orbit in $\mathbb W_{p,q}$.

\begin{thm} [\cite{Jen-09}] Suppose $1\le  p < q$ are coprime integers. For $w\in \mathbb W_{p,q}$, the product $B(w)$
is maximized precisely when w is balanced.
\end{thm}

For instance, put $p=2, q=5$. Here there are only two possible orbits, namely, those of $w=10100$ and $v=11000$. We have $B(w)=b(10100)\times b(01001)\times b(10010)\times b(00101)\times b(01010)=20\times 9\times 18 \times 5\times 10=162000$, whereas $B(v)=b(11000)\times b(10001)\times b(00011)\times b(00110)\times b(01100)=24\times 17\times 3\times 6\times 12=88128$.

\section{Maximizing measures}

Let $T:[0,1)\to [0,1)$ and denote the space of all $T$-invariant measures by $\M(T)$. We say that a measure $\mu$ is {\em majorated} by $\nu$ (notation: $\mu\prec\nu$) if $\int_0^1 f\ d\mu\le \int_0^1 f\ d\nu$ for any convex function $f:[0,1)\to [0,1)$. Put for any $\ga\in(0,1)$,
\[
\M_\ga=\{\mu\in \M(T) : \text{bar}(\mu)=\ga\},
\]
where
\[
\text{bar}(\mu)=\int_0^1 x\ d\mu(x),
\]
i.e., the {\em barycentre} of $\mu$.

\begin{thm} [\cite{Jen-08}] Let $Tx=2x\bmod1$. For any $\ga\in[0,1]$, the partially ordered set $(\M_\ga, \prec)$ has a least element. This least element is the Sturmian measure $S_\ga$ of rotation number $\ga$.
\end{thm}

Here $S_\ga$ is the following. Let $\varphi:[0,1)\to [0,1)$ be defined as follows:
\[
\varphi_\ga(x)=\sum_{n=0}^\infty \frac{\chi_{[1-\ga,1)}(x+n\ga\bmod1)}{2^{n+1}},
\]
i.e., the binary sum of the standard symbolic sequence associated with the rotation by $\ga$ with a starting point $x$. Then the Sturmian measure $S_\ga$ is the push forward of the Lebesgue measure on $[0,1)$ under $\varphi_\ga$.

If $\ga$ is rational, then $S_\ga$ sits on a finite set. For instance, the support of $S_{2/5}$ is the orbit of the binary sum of $00101\ 00101\ 00101\dots$ under $T$, i.e., the set $\left\{\frac5{31}, \frac{10}{31}, \frac{20}{31}, \frac{9}{31}, \frac{18}{31}\right\}$, each point having the $S_{2/5}$-measure of $1/5$.

If $\ga$ is irrational, then $S_\ga$ is supported by a Cantor set. For instance, if $\ga=\frac{3-\sqrt5}2$, then $\text{supp}(S_\ga)$ is the closure of the set of all shifts of the Fibonacci word $f=0010100100101\dots$ (It is indeed a Cantor set because $f$ has such a low complexity!)

There are other papers in this area which produce Sturmian sequences in similar optimization problems. For instance, one may replace the class of convex functions with increasing functions and consider the {\em $\beta$-transformation} given by $\tau_\be x=\be x\bmod1$ (with $\be>1$) and the $\be$-shift -- the subshift on the alphabet $\{0,1,\dots,\lceil \be \rceil-1\}$ which corresponds to the natural partition $[0,1)=[0,\beta^{-1})\cup [\beta^{-1},2\beta^{-1})\cup\dots \cup [(\lceil\beta\rceil-1) \beta^{-1},1)$ for $\tau_\be$.

It has been shown in \cite{AJ} that the $\be$-shift has a largest shift-invariant measure if and only if $\be$ is an algebraic integer of a special form. (In particular, if $1<\be<2$, then it has to be {\em multinacci}, i.e., the dominant root of $x^m=x^{m-1}+\dots+x+1$ for some $m\ge2$.) In this case the largest shift-invariant measure on the $\be$-subshift is the
unique one supported by the periodic shift-orbit generated by its lexicographically largest element. The supporting measure is always Sturmian.

Another direction in this line of research is concerned with imposing no extra conditions of the class of functions but instead considering a specific (usually, one-parameter) family of those. For example, let, as above, $T$ be the doubling map, and let $g_\theta(x)=\cos 2\pi(x-\theta)$ or $f_\theta(x)=1-4\text{dist}_{\mathbb T}(x,\theta)$, where $\text{dist}_{\mathbb T}$ is the distance on the circle $\mathbb R/\mathbb Z$. In both cases maximizing measures are Sturmian -- see \cite{B, ADJR} and references therein.

Thus, in questions concerning maximizing measures, the Sturmian measures seem to be a very robust class.

\section{Tetris heaps}

\begin{figure}
\centering \scalebox{0.8} {\includegraphics{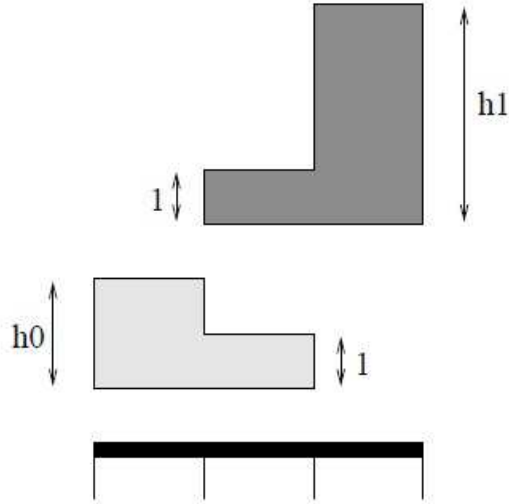}}
\caption{Tetris heaps}\label{fig:1}
\end{figure}

Consider a version of popular Tetris game with two pieces, 0 and 1 -- see Fig.~\ref{fig:1}. We will be interested in stacking these pieces in such a way that the height of the heap is minimal.

More precisely, for a finite 0-1 word $w$ we define $h(w)$ to be the height of the heap specified by $w$ and put
\[
\rho_{\text{min}} = \liminf_{n\to\infty} \min_{w\in\{0,1\}^n} \frac {h(w)}n.
\]
An {\em optimal schedule} is an infinite 0-1 word $u$ such that
\[
\lim_{n\to\infty} \frac{h(u[n])}n = \rho_{\text{min}},
\]
where $u[n]$ is the prefix of length $n$ of $u$.

\begin{thm} [\cite{MV}] Let us consider a heap model with two pieces. There exists an optimal schedule which is balanced -- either periodic or Sturmian.
\end{thm}

The authors of \cite{MV} characterize the cases where the optimal is periodic and the ones where it is Sturmian. The proof is constructive, providing an explicit optimal schedule. For a more general approach (using a special class of iterated function systems) see \cite{BM}.

\section{Joint spectral radius}

Given a finite set of $d \times d$ real matrices $\mathcal{A} = \{A_0,\ldots,A_{r-1}\}$, we define the {\em joint spectral radius} $\varrho(\mathcal{A})$ to be the quantity
\[\varrho(\mathcal{A}):=\limsup_{n \to \infty} \max\left\{\left\|A_{i_1}\cdots A_{i_n}\right\|^{1/n} \colon i_j \in \{0,\ldots,r-1\}\right\},\]
a definition introduced by G.-C.~Rota and G.~Strang in 1960.

A $\{0,\dots,r-1\}$-sequence $(i_1,i_2,\dots)$ is called {\em maximizing} if
\[
\lim_{n \to \infty} \|A_{i_1}\cdots A_{i_n}\|^{1/n} =\varrho(\mathcal{A}).
\]
This is somewhat similar to the Tetris model considered above, which is reflected in the general model accounting for both set-ups considered in \cite{BM}.

\begin{example}Put $\mathcal A=\{A_0,A_1\}$, where
\[
A_0=\begin{pmatrix} 1 & 1 \\ 0 & 1 \end{pmatrix}, \ A_1=\begin{pmatrix} 1 & 0 \\ 1 & 1 \end{pmatrix}.
\]
Then it turns out that $010101\dots$ is maximizing and furthermore, any maximizing sequence for $\mathcal A$ has the same growth rate of the corresponding sequence of matrix products. Consequently, $\varrho(\mathcal A)=(\rho(A_0A_1))^{1/2}=(1+\sqrt5)/2$.
\end{example}

Now consider the one-parameter family of pairs $\mathcal{A}_\alpha:=\{A_0, \al A_1\}$ with $\al\in[0,1]$. Is it true that for any fixed $\alpha$ any maximizing sequence is ``essentially periodic'' like in the case $\alpha=1$?

More precisely, a set of matrices $\mathcal A$ is said to have the {\em finiteness property} if there exists an eventually periodic maximizing sequence for $\mathcal A$. It was shown in the PhD thesis \cite{Th} that if $\al\in[4/5,1]$, then we have the same conclusion as for $\alpha=1$. What about the case $\alpha<4/5?$

We say that an infinite 0-1 word $w$ is {\em recurrent} if any of its subwords occurs in $w$ infinitely often. For each $\gamma \in [0,1]$, let $X_\gamma$ denote the set of all recurrent balanced infinite words whose 1-ratio is equal to $\gamma$. For any rational $\ga$ the set $X_\ga$ is finite and it is a continuum (called a {\em Sturmian system}) for any irrational $\ga$.

We say that $\ga$ is an {\em optimal 1-ratio} for $\mathcal A_\al$ if there exists a maximizing sequence for $\mathcal A_\al$ with the 1-ratio $\ga$.

\begin{thm}[\cite{hmst}]
There exists a continuous, non-decreasing surjection $\mathfrak{r} : [0,1] \to [0,\frac12]$ such that for each $\alpha$, $\ga=\mathfrak{r}(\alpha)$ is the unique optimal $1$-ratio of $\mathcal{A}_\alpha$.

Furthermore, for each $\alpha \in [0,1]$, every element of $X_\ga$ is a maximizing sequence.
\end{thm}

Thus, if one takes any irrational $\ga\in(0,1/2)$, then there exists (in fact, unique -- see \cite{MS}) $\al\in[0,1]$ such that any maximizing sequence for $\mathcal A_\al$ has the 1-ratio $\ga$, i.e., cannot be periodic. This disproves the {\em Finiteness Conjecture} which asserts that any maximizing sequence for an arbitrary set of matrices should be periodic. (The first counterexample to the FC appeared in \cite{BM}, however it was not explicit.)

Moreover, there is an explicit formula for such an $\alpha$ in terms of the elements of the continued fraction expansion of $\ga$. Namely, let $$\gamma=[a_1,a_2,\dots]$$ denote the continued fraction expansion of $\ga$ with $p_n/q_n$ being the $n$th convergent. Recall that the sequence of {\em standard words} specified by $\gamma$ is given by $s_{-1}=1, s_0=0,
s_{n+1}=s_n^{a_{n+1}}s_{n-1}, \ n\ge0$. It is obvious that $s_n$ is a prefix of $s_{n+1}$ and that the length of $s_n$ tends to the infinity. It is also well known that the 1-ratio of $s_n$ is $p_n/q_n$ (see \cite{Loth}).

Put $s_\infty=\lim_{n\to\infty}s_n$ (its 1-ratio is thus $\ga$) and define the sequence of $2\times 2$ matrices as follows: $B_{-1}=A_1, B_0=A_0$, and
\[
B_{n+1}=B_n^{a_{n+1}}B_{n-1},\quad n\ge0.
\]
Put $\rho_n=\rho(B_n)$ (the spectral radius of $B_n$) and $\tau_n=\text{tr}(B_n)$.

\begin{thm} [\cite{hmst, MS}]
If $\mathfrak r^{-1}(\gamma)=\{\alpha\}$, we have
\[
\alpha =\lim_{n\to\infty}\left(\frac{\rho_n^{q_{n+1}}}
{\rho_{n+1}^{q_n}}\right)^{(-1)^n} \\
 = \prod_{n=0}^\infty\left(\frac{\rho_n^{d_{n+1}}\rho_{n-1}}{\rho_{n+1}}
\right)^{(-1)^nq_n}.
\]
\end{thm}

In particular, if $\gamma_*=\frac{3-\sqrt5}2$ (i.e., $d_n\equiv1$ for all $n$), then (see \cite{hmst})
\[
\alpha_*:=\lim_{n \to \infty}
\left(\frac{\tau_n^{F_{n+1}}}{\tau_{n+1}^{F_n}}\right)^{(-1)^n}= \prod_{n=1}^\infty \left(1-\frac{\tau_{n-1}}{\tau_n \tau_{n+1}}\right)^{(-1)^n F_{n+1}},
\]
where $F_0:=0, F_1:=1$ and $F_{n+1}:=F_n+F_{n-1}$ is the Fibonacci sequence and, as it turns out, $\tau_0:=1$, $\tau_1,\tau_2:=2$ and $\tau_{n+1}:=\tau_n\tau_{n-1}-\tau_{n-2}$. The infinite product converges superexponentially fast, and
\[\alpha_* \simeq 0.749326546330367557943961948091344672091327
\ldots
\]
Similarly, one can easily compute $\mathfrak r^{-1}(\gamma)$ for any irrational $\gamma$ with a very high precision. For more detail see \cite{MS}.


\section{One-dimensional Wigner lattices}

Consider the one-dimensional lattice $\mathbb Z_+$ whose each node is either occupied by an electron or is empty (``occupied by a hole''). The electrons' interaction is given by a potential $V$ which is assumed to be convex and vanish at the infinity, which is a rather weak assumption (such is, for instance, a Coulomb potential). More precisely, the energy of the system is given by
\[
E=\frac12\sum_{x_i,x_j}V(|x_i-x_j|).
\]
Let $\gamma\in\mathbb Q$ denote the (fixed) ratio of electrons on the lattice.

Then, as noticed by Hubbard \cite{Hu}, the lowest-energy configuration with respect to this potential is attained at balanced words (see Table~II from the cited paper). It should be mentioned that the author does not use this terminology and provides a proof of his claim only for certain special cases of $\gamma$.

The same result has been rediscovered by several authors for various physical models -- see, e.g., \cite{Pu}.

Summing up, there are numerous -- seemingly unrelated -- areas of mathematics and physics in which optimization problems yield balanced words. Gaining a better understanding of this phenomenon looks like a perspective line of research.

\bibliographystyle{eptcs}

\begin{thebibliography}{99}

\bibitem{ADJR} V. Anagnostopoulou, K. Diaz-Ordaz, O. Jenkinson and C. Richard, {\em Sturmian maximizing measures for the
piecewise-linear cosine family}, preprint, see http://www.maths.qmul.ac.uk/\~{}omj/.

\bibitem{AJ} V. Anagnostopoulou and O. Jenkinson, {\em Which beta-shifts have a largest invariant measure?} J. London Math. Soc. {\bf 79} (2009), 445--464. \doi{10.1112/jlms/jdn070}

\bibitem{B}T. Bousch, {\em Le poisson n'a pas d'ar\^{e}tes},  Ann. Inst. H. Poincar\'e Probab. Statist. {\bf 26} (2000), 489-–508. \doi{10.1016/S0246-0203(00)00132-1}

\bibitem{BM}T. Bousch and J. Mairesse, {\em Asymptotic height optimization for topical  IFS, Tetris  heaps, and the finiteness conjecture}, J. Amer. Math. Soc. {\bf 15} (2002), 77--111. \doi{10.1090/S0894-0347-01-00378-2}

\bibitem{Ha}B. Hajek, {\em Extremal splittings of point processes}, Math. Oper. Res. {\bf 10} (1985), 543--556. \doi{10.1287/moor.10.4.543}

\bibitem{Hu} J. Hubbard, {\em Generalized Wigner lattices in one dimension and some applications to tetracyanoquinodimethane (TCNQ) salts},
Phys. Rev. B {\bf 17} (1978), 494--505. \doi{10.1103/PhysRevB.17.494}

\bibitem{Jen-08} O. Jenkinson, {\em A partial order on $\times$2-invariant measures}, Math. Res. Lett. {\bf 15} (2008), 893--900.
    
\bibitem{Jen-09}O. Jenkinson, {\em Balanced words and majorization}, Discr. Math. Alg. Appl. {\bf 1} (2009), 485--498. \doi{10.1142/S179383090900035X}

\bibitem{hmst} K. Hare, I. Morris, N. Sidorov and J. Theys, {\em An explicit counterexample to the Lagarias-Wang finiteness conjecture}, Adv. Math. {\bf 226} (2011), 4667--4701. \doi{
10.1016/j.aim.2010.12.012}

\bibitem{Loth}M. Lothaire, {\em Algebraic combinatorics on words, Encyclopedia of Mathematics and its Applications}, Cambridge
University Press, Cambridge, 2002. \doi{10.1145/1412700.1412706}

\bibitem{MV}J. Mairesse and L. Vuillon, {\em Asymptotic behavior in a heap model with two pieces}, Theoret. Comp. Sci. {\bf 270} (2002), 525–-560. \doi{10.1016/S0304-3975(01)00004-4}

\bibitem{MS}I. Morris and N. Sidorov, {\em A Devil's staircase associated to the joint spectral radius of a pair of matrices}, in preparation.

\bibitem{Pu} V. L. Pokrovsky and G. V. Uimin, {\em On the properties of monolayers of adsorbed atoms}, J. Phys. C {\bf 11} (1978), 3535--3549. \doi{10.1088/0022-3719/11/16/022}

\bibitem{Th}J. Theys, {\em Joint Spectral Radius: theory and approximations}, PhD thesis, Universit\'e Catholique de Louvain, 2005.

\end{thebibliography}

\end{document}